\documentclass[12pt]{iopart}

\usepackage[dvips]{graphicx}
\usepackage{dcolumn}	
\usepackage{bm}		
 \usepackage{latexsym}
\usepackage{iopams}
\usepackage{color}

 \usepackage{dsfont}				
\usepackage{psfrag}
\usepackage{pstricks,pst-grad}

 \newcommand{\new}[1]{#1}
 \newcommand{\old}[1]{}

\newcommand{\im}{i}

\newcommand{\nsi}{_{n\sigma}}
\newcommand{\nsp}{n'\sigma '}

\newcommand{\ket}[1]{\;|{#1}\rangle}
\newcommand{\bra}[1]{\langle {#1}|\;}
\newcommand{\be}{\begin{equation}}
\newcommand{\ee}{\end{equation}}
\newcommand{\bdm}{\begin{displaymath}}
\newcommand{\edm}{\end{displaymath}}
\newcommand{\bea}{\begin{eqnarray}}
\newcommand{\eea}{\end{eqnarray}}
\newcommand{\beaa}{\begin{eqnarray*}}
\newcommand{\eeaa}{\end{eqnarray*}}
\newcommand{\arr}[1]{\left(\begin{array}{c} #1 \end{array}\right)}
\newcommand{\Ham}{\hat{\mathcal{H}}}

\newcommand{\Tchar}{\Delta T(E;U_0)}
\newcommand{\Tspini}{\Delta T_{\mathrm{S}}(E;U_0)}

\newcommand{\UB}{U_{\mathrm{B}}}
\newcommand{\Fermi}{\varepsilon _\mathrm{F}}
\newcommand{\BZ}[1]{(\ref{#1})}

\eqnobysec

\begin{document}

\title[Zeeman ratchets: pure spin current generation in mesoscopic conductors...]{Zeeman ratchets: pure spin current generation in mesoscopic conductors with non-uniform magnetic fields}

\author{Matthias Scheid$^1$, Dario Bercioux$^{1,2}$, and Klaus Richter$^1$}

\address{$^1$ Institut f\"ur Theoretische Physik, Universit\"at Regensburg, D-93040 Regensburg, Germany}
\address{$^2$ Physikalisches Institut, Albert-Ludwigs-Universit\"at, D-79104 Freiburg, Germany}
\eads{\mailto{Matthias.Scheid@physik.uni-regensburg.de}, \mailto{Dario.Bercioux@physik.uni-freiburg.de}}

\begin{abstract}
We consider the possibility to employ a quantum wire realized in 
a two-dimensional electron gas (2DEG) as a spin ratchet. We show that
a net spin current without accompanying net charge transport can be induced 
in the nonlinear regime by an unbiased external driving via an ac voltage 
applied between the contacts at the ends of the quantum wire. To achieve 
this we make use of the coupling of the electron spin to inhomogenous magnetic 
fields created by ferromagnetic stripes patterned on the semiconductor 
heterostructure that harbours the 2DEG. Using recursive Green function
techniques we numerically study two different setups, consisting of one 
and two ferromagnetic stripes, respectively.
\end{abstract}
\submitto{\NJP}
\maketitle

\section{Introduction}
\emph{Spintronics} as an emerging field of physics has attracted considerable 
attention in recent years and has developed into various inter-related branches
covered in this topical issue. Spintronics is devoted to employ the spin degree of
freedom for information storage and as another means for extending
the functionality of electronic systems. Semiconductor spintronics, as one subfield,
is guided by the idea of combining concepts of spin electronics with the 
established techniques and advantages of semiconductor physics and nanostructures.
Up to now, their properties used mainly rely on the charge 
degree of freedom alone.
Many ideas for employing spin-polarized currents have been put forward since the 
seminal propsal by Datta and Das for a spin transistor~\cite{datta:1990} based 
on spin precession controlled by an external electric field through spin-orbit (SO)
coupling~\cite{rashba:1960}. These proposals usually require spin injection, 
more generally, the creation of spin-polarized particles in these
materials. Spin injection from a ferromagnetic metal source into semiconductors
is hindered by a fundamental obstacle originating from the conductivity mismatch 
between these materials~\cite{schmidt:2000}. Though this problem may be partially
circumvented, for instance at low temperatures by tailoring 
dilute-magnetic-semiconductor/semiconductor interfaces~\cite{fiederling:1999,ohno:1999}
enabling considerable spin-polarization ratios of the order of 
$90\%$~\cite{fiederling:1999}, building all-semiconductor 
sources of spin-polarized electrons is still a challenge.
 
Alternatively, several techniques to intrinsically create spin currents in 
non-magnetic systems have been put forward:
Non-equilibrium spin-polarized currents have been created in two-dimensional electron 
gases (2DEGs) realized in zinc-blende-based heterostructures by means of optical 
pumping~\cite{ganichev:2001}. The irradiation of the 2DEG with circularly polarized 
light results in a spin photocurrent caused by the non-uniform distribution of the 
photoexcited carriers in \textbf{k}-space owing to optical selection rules and energy 
and momentum conservation.
The recently proposed intrinsic spin-Hall effect in 
$p$-doped~\cite{murakami:2003} bulk systems and in 2DEGs~\cite{sinova:2003} offers 
the principle possibility for spin current generation and manipulation in high 
mobility semiconducting systems. 
The combined effect of an applied electric field and the torque induced by SO coupling 
tilts the spins out of the 2DEG plane. This leads to spin accumulation at both sides
of the sample in the direction perpendicular to the applied electric field.
Spin-Hall effects have been observed experimentally both in semiconducting systems
by optical detection techniques~\cite{kato:2004,wunderlich:2005,sih:2005} 
and in metallic systems with an all-electrical setup~\cite{valenzuela:2006,kimura:2007}.

In the context of mesoscopic physics, further concepts such as 
adiabatic spin pumping~\cite{sharma:2001,mucciolo:2002,governale:2003} and coherent spin 
ratchets~\cite{pfund:2006,Proc} have been proposed for generating spin-polarized
currents. This can be achieved by exploiting the magnetic properties of the 
semiconducting material, \emph{i.e.} intrinsic spin-orbit interaction or the Zeeman coupling 
to external magnetic fields.

Adiabatic quantum pumping involves the generation of a directed current in the absence of a 
bias voltage by periodic modulations of two or more system parameters, such as,
\emph{e.g.}, the shape of the system or a magnetic field~\cite{brouwer:1998}. The spin 
analog of the charge quantum pump can be achieved through an external magnetic 
field~\cite{mucciolo:2002} or spin-orbit interaction~\cite{governale:2003} 
in order to filter the pumped current. 
For the case of an external magnetic field, it has been experimentally
confirmed that under specific circumstances a spin current can be extracted from
spin-dependent conductance fluctuations without accompanying net charge 
flow~\cite{watson:2003}.

{\em Ratchets}~\cite{reimann:2002} are generally systems with broken inversion 
(left/right) symmetry 
that generate directed net (particle) currents upon external AC driving in the
absence of a net (time-averaged) bias potential. Thereby they have much in
common with current rectifiers, though there are differences, in particular
in the dissipative case~\cite{reimann:2002}. The theoretical concept of ratchets, 
originally introduced for classical dynamics, was later extended to the quantum 
dissipative regime~\cite{reimann:1997}. 
As a main feature, which distinguishes them also from rectfiers in the
usual sense, quantum ratchets exhibit current reversal upon changing, e.g.,
the temperature or energy.
Such quantum ratchets were experimentally realized in semiconductor 
heterostructures by demonstrating directed charge flow in a chain of asymmetric 
ballistic electron cavities in 
the low-temperature regime, where the dynamics was close to coherent~\cite{linke:1999}.
In this and further experimental investigations~\cite{linke:2002} 
also the charge current reversal phenomenon has been demonstrated.

Very recently, we have proposed the generalization of the ratchet mechanism,
extensively explored for particle motion, to generate directed spin currents. 
Spin ratchets
require a coupling to the electron spin which, \emph{e.g.}, can be provided via
spin-orbit interaction or external (non-uniform) magnetic fields. In Ref.~\cite{pfund:2006}
{\em spin-orbit ratchets} have been considered in the coherent regime. There a proof
of principle for a net ratchet spin current (in absence of an average charge current)
has been given and confirmed by numerical calculations for experimentally
accessible parameters for GaAs-based heterostructures.

In the present paper we further investigate the possibility of employing {\em Zeeman
ratchets} for spin current generation, \emph{i.e.} by considering mesoscopic conductors
with a spatially varying Zeeman term and subject to an AC bias. In the simplest 
case of one-dimensional motion, and assuming preserved spin states, spin-up and 
-down electrons will experience a Zeeman term, $\pm (g^*/2)\mu_{\rm B} B(x)$,
where $B(x)$ is the external non-uniform magnetic field, $g^*$ the effective
gyroscopic factor and $\mu_{\rm B}$ the Bohr magneton. If $B(x) \neq B(-x)$,
then the different spin species experience opposite asymmetric Zeeman ratchet potentials.
In close analogy to the particle ratchet mechanism described above, the different
spins are expected to be predominantly driven into opposite directions upon external 
driving, resulting in a spin-polarized current. This mechanism has been studied 
and confirmed in Ref.~\cite{Proc}. 

In the present work we relax both assumptions of one-dimensional motion
\cite{braunecker:2006} and conserved
spin directions, \emph{e.g.} the notion of two independent spin species, and explicitly
include spin-flip effects. Moreover, we work out how such spin flip processes
can be invoked to engineer and tune ratchet spin conductances.
To this end we consider spin ratchet effects of a system consisting of a 
two-dimensional quantum wire embedded in 2DEG and subject to an AC bias in between 
two ohmic contacts.
The non-uniform $B$-field is created by the magnetic fringe fields of 
ferromagnetic stripes patterned on the semiconductor heterostructure~\cite{PeetersPionier}.

%
%
%
\begin{figure*}
	\begin{center}
	\includegraphics[width=\textwidth]{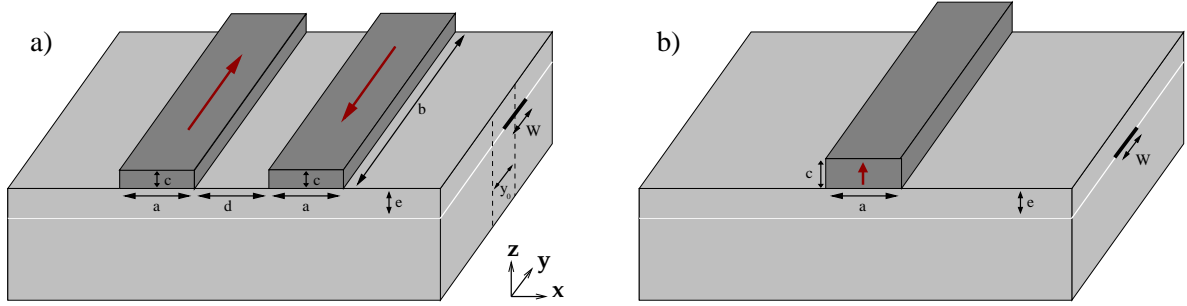}
	\caption{\label{figure:1}Ferromagnetic stripes (magnetization direction
	given by red arrows) on top of a semiconductor heterostructure that harbors a
	two-dimensional electron gas (indicated by white lines) containing a quantum
	wire (black) of width $W$. 
    a) \old{Configurations}\new{Setup A}: two stripes with antiparallel in-plane
        magnetization, see Sec.~\ref{S3}; b) \new{setup B}: one stripe with 
	out-of-plane  magnetization, see Sec.~\ref{S4}.}
	\end{center}
    \end{figure*}
%
%

This paper is organized as follows: In Sec.~\ref{K9} we introduce the model for the spin ratchet. 
We thereby specify the driving of the ratchet and the evaluation of the net charge and spin currents. 
In Sec.~\ref{S3} we study transport through the quantum wire subject to the 
magnetic fringe fields of two ferromagnetic stripes with antiparallel magnetization
perpendicular to the quantum wire in the plane of the 2DEG. 
This configuration (setup A in Fig.~1a) allows us to study the transition from 
decoupled to strongly coupled spin states and its implications on transport and 
thus also on the ratchet currents. In Sec.~\ref{S4} we then investigate the
conductor subject to the fringe field of a single ferromagnetic stripe magnetized 
perpendicular to the plane of the 2DEG (setup B in Fig.~1b)). 
Using symmetry arguments and numerical calculations we demonstrate that
the two setups introduced in Sec.~\ref{S3} and~\ref{S4} act as  spin ratchets. 
After summarizing in Sec.~\ref{Conclusions}, we close the paper with an appendix 
on the general derivation of an expression for the spin current within the framework 
of the multi-terminal Landauer-B\"uttiker formalism and a second appendix
including the derivation of symmetry relations for the transport properties 
at finite applied bias.

\section{Model and formalism}\label{K9}

We consider a quantum wire in the $x$-direction embedded in a 2DEG in the ($x,y$) plane.
The system is subject to a non-uniform magnetic field $\vec{B}(x,y)$, due to the fringe fields of ferromagnetic 
stripes patterned on top of the 2DEG (see Fig.~\ref{figure:1}).
Their deposition on a semiconductor heterostructure can be accomplished with electron 
beam lithography and lift-off techniques~\cite{WeissOszis,Kubrak}.
Near-surface 2DEGs can be fabricated to lie only a few tens of nanometers beneath 
the surface~\cite{Kubrak}, thereby achieving magnetic field values of up to $0.5$T 
with thin ferromagnetic films~\cite{Kubrak2}. The magnetic fringe field of a 
ferromagnet with homogeneous magnetization $\vec{M}$ is given by~\cite{Jackson}
%
%
\begin{equation}
\vec{B}(\vec{r})=-\frac{\mu _0}{4\pi}\vec{\nabla}
\oint _S \mathrm{d}a' \;\frac{\vec{M}\cdot\hat{u}(\vec{r}^{\;\prime} )}{|\vec{r}-\vec{r}^{\;\prime} |},
\label{Bfin}
\end{equation}
%
%
where the integration runs over the surface $S$ of the ferromagnetic stripe, and $\hat{u}(\vec{r}^{\;\prime} )$ 
is the unit vector normal to the surface of the stripe at position $\vec{r}^{\;\prime}$. 
Accordingly, the corresponding vector potential in the Coulomb gauge,
$\vec{\nabla}\cdot\vec{A}=0$, is given by~\cite{Jackson}
%
%
\be
\vec{A}(\vec{r})=\frac{\mu _0}{4\pi}
\oint _S \mathrm{d}a' \frac{\vec{M}\times\hat{u}(\vec{r}^{\;\prime} )}{|\vec{r}-\vec{r}^{\;\prime}|}.
\label{Afin}
\ee
%
%
We model the wire, where electron transport is assumed to be phase coherent, 
by the single-particle Hamiltonian
%
%
\begin{equation}\label{Hsim}
\hat{\mathcal{H}}_0=\frac{\Pi _x(x,y)^2 +\Pi _y(x,y)^2}
{2m^*}+ \frac{g^*\mu_\mathrm{B}}{2} \vec{B}(x,y)\cdot\vec{\sigma }+V(y),
\end{equation}
%
%
where $g^*$ is the effective gyroscopic factor, $m^*$ the effective electron mass, $\mu_\mathrm{B}$ the Bohr magneton and $\vec{\sigma}$ the vector of the Pauli matrices. 
The term $V(y)$ denotes the lateral confining potential defining the quantum wire. 
Orbital effects due to the magnetic field are accounted for by the vector potential $\vec{A}(x,y)$ in $\vec{\Pi}(x,y)=\vec{p}-e\vec{A}(x,y)$. 
Spin effects in transport through the wire enter via the Zeeman term 
$(g^*\mu_\mathrm{B}/2) \vec{B}\cdot\vec{\sigma }$ coupling the spin degree of freedom to the external magnetic field. 
For a proper treatment of the spin evolution, the inclusion of the full magnetic 
field profile is mandatory~\cite{FengZhaiStripe}; disregarding~\cite{Maowang}
one of the magnetic field components $B_i$ may lead to an oversimplification
of the problem.

To obtain a significant spin ratchet effect materials with a large 
$g^*$ factor are most suitable. In this respect dilute magnetic semiconductors (DMS) 
represent a promising class of materials. Recent measurements have shown values of 
$g^*>100$~\cite{DMSDiodeMolenkamp} where in addition to the intrinsic 
$g$-factor, an additional contribution to $g^*$ appears, owing to exchange coupling 
among the electron spins and the magnetic ions present in the DMS~\cite{FurdDMSRev}.
These materials with a large $g^*$ factor can also exhibit 
large SO coupling values. The working principle of a spin ratchet based on SO coupling 
has been already investigated in Ref.~\cite{pfund:2006}. As one result of this study 
a spin-ratchet effect can only occur if there exists the possibility to mix 
different transverse channels of the wire. This can only happen when more than one 
open channel is taking part in the transport and electrostatic barriers induce this 
mixing of different bands.  As we will show below, the spin-ratchet effect created
by the setup presented in this paper does not rely on this condition and is already 
present when only the first conducting channel is opened. Due to this fact and the 
absence of any electrostatic barriers responsible for the mixing of different subbands 
we disregard here the SO coupling and focus on the effects due to the presence of 
the magnetic stripes.

In the present work where we consider disorder-free ballistic motion, we refer to
nonmagnetic high-mobility semiconductors. To be definite we chose throughout the
paper parameters for InAs 2DEGs with typical values of $m^*=0.024m_0$\cite{SemiProp}, 
where $m_0$ is the free electron mass, and $|g^*|=15$. InAs is well suited due to its large
$g^*$ factor and the property that InAs 2DEGs can be fabricated close to the
surface where the magnetic stripes are located. We
assume that the stripes possess a magnetization $\mu _0M=3$T.

The charge current $I_{\rm C}$ through the wire is evaluated within 
the Landauer approach. 
For coherent transport in a two-terminal device the current can be expressed
as
%
%
\begin{equation}\label{Ic}
I_{\mathrm{C}}=
-\frac{e}{h}\int_{0}^{\infty} \mathrm{d}E \;\big[ f(E;\mu _\mathrm{L}) - f(E;\mu _\mathrm{R} ) \big] \;T(E)
\end{equation}
%
%
in terms of the quantum  probability $T(E)$ for electrons with energy $E$ to be 
transmitted from the lead at higher to the  lead at lower potential. In Eq.~(\ref{Ic}),
$f(E;\mu_\mathrm{L/R})$ is the Fermi function for the left/right lead with chemical 
potential $\mu _\mathrm{L/R}$.
 
The spin current $I_{\mathrm{S}}(x)$ passing
a cross section ($x=\,$const) is given, for a wavefunction
$\Psi(x,y)$, by
%
%
\bdm
I_{\mathrm{S}}(x)=\int \mathrm{d}y \Psi ^*(x,y)\hat{J}_{\mathrm{S}}\Psi(x,y) \, .
\edm
%
%
Here we use the most common definition~\cite{RashbaSpinDyn} of the spin current
operator $\hat{J}_{\mathrm{S}}$  which, with respect to an arbitrary 
quantization axis $\hat{u}$, reads
%
%
\be
\label{eq:def-spin-current}
\hat{J}_{\mathrm{S}}=\frac{\hbar}{2}\frac{\hbar}{2m^*i}\left(\vec{\sigma}\cdot\hat{u}\right)\left(
\overrightarrow{\frac{\partial}{\partial x}}
- \overleftarrow{\frac{\partial}{\partial x}} \right)
\ee
%
%
inside the leads. The partial derivatives in 
(\ref{eq:def-spin-current}) act on expressions to their right and left 
(indicated by the arrows).

Contrary to the charge current that obeys a continuity equation, 
the spin current can take different values if evaluated in the left or in the right lead. 
This usually happens in systems where the Hamiltonian does not commute with the Pauli 
matrices $\vec{\sigma}$ giving rise to a torque inside 
the scattering region~\cite{waintal:2000}, which can change the spin state of the 
electrons. For this reason we will explicitly label the lead, where we compute the spin current. 
Although there is some freedom in the choice of the spin current 
operator~\cite{shi:2006}, here we evaluate the spin current
inside the magnetic field free leads, where a tourque term in the 
continuity equation for spin is absent. Therefore the spin current 
inside the leads is a well defined quantity, which can be measured 
in principle.
As derived in~\ref{App} (Eq.~\ref{spincurr}) the corresponding spin current  
in the right lead reads
%
%
\begin{equation}\label{Is}
I_{\mathrm{S}}=
\frac{1}{4\pi}\int_{0}^{\infty} \!\!\mathrm{d}E\;
\big[ f(E;\mu _\mathrm{L} ) - f(E;\mu _\mathrm{R})  \big] T_{\mathrm{S}}(E) \, ,
\end{equation}
%
%
with the spin transmission probability defined as
%
%
\be
T_\mathrm{S}(E)=\sum_{\sigma =\pm 1}\big[ T_{+,\sigma}(E) -T_{-,\sigma}(E) \big] \, .
\label{spintrans}
\ee
%
%
Here $T_{\sigma ',\sigma}$ is the probability for an electron with initial 
spin state $\sigma$ to be transmitted from the left lead into the spin state 
$\sigma '$ inside the right lead, see Eq. (\ref{TransDef}) in~\ref{App}. 
To obtain the corresponding spin current in the left lead one has to replace 
$T_{\sigma', \sigma}(E)$ in Eq.~(\ref{spintrans}) 
by the corresponding probabilities
$T'_{\sigma ', \sigma}(E)$ 
for transmission from the right to the left lead.

Ordinary particle ratchets give rise to a net drift motion of particles in one 
preferential direction upon AC driving without net bias (rocking ratchet).
Below we will generalize this concept to induce spin-dependent ratchet currents 
correspondingly. The AC driving can be 
considered as adiabatic, since the timescales for the variation of an external bias 
potential are long compared to the relevant time scales for charge transmission through 
the device. For a proof of principle we assume an adiabatic unbiased square-wave driving 
with period $t_0$. 
The system is periodically switched between two rocking conditions, labeled by bias
$\pm U_0$ ($U_0>0$). The electro-chemical potential $\mu _{\mathrm{L/R}}$ of 
the left/right reservoir is changed periodically in time according to
%
\begin{equation}
\mu _{\mathrm{L/R}}(t)=
\cases{\Fermi \pm U_0 / 2 \qquad\mathrm{for}\quad 0\leq t<t_0 / 2    \\
\Fermi \mp U_0 / 2 \qquad\mathrm{for}\quad t_0/ 2\leq t< t_0}\, ,
\label{adrocking}
\end{equation}
%
\emph{i.e.} $\mu _{\mathrm{L}/\mathrm{R}}(t)=\mu _{\mathrm{L}/\mathrm{R}}(t+t_0)$.
In the adiabatic limit 
considered, the system is assumed to be in a steady state between the switching events. 
Then the ratchet charge and spin currents inside the wire are obtained upon averaging 
Eqs.~(\ref{Ic},\ref{Is}) between the two rocking situations
%
%
\numparts
\begin{eqnarray}\label{Icnet}
\langle I_{\mathrm{C}}(\Fermi ,U_0)\rangle 
& = & \frac{1}{2}\left[I_{\mathrm{C}}(\Fermi ,+U_0)+I_{\mathrm{C}}(\Fermi ,-U_0)\right]\\
 & = &\! -\frac{e}{2h}\int_{0}^{\infty} \!\! \mathrm{d}E \;\Delta f(E;\Fermi ,U_0) 
 \Tchar  \, ,
 \nonumber
\end{eqnarray}
\begin{eqnarray}\label{Isnet}
\langle I_{\mathrm{S}}(\Fermi ,U_0)\rangle & = & 
\frac{1}{2}\left[I_{\mathrm{S}}(\Fermi ,+U_0)+ I_{\mathrm{S}}(\Fermi ,-U_0)\right]\\
& = & \!\frac{1}{8\pi}\int_{0}^{\infty}\!\! \mathrm{d}E \;\Delta f(E;\Fermi ,U_0)
\Tspini \, , \nonumber
\end{eqnarray}
\endnumparts
%
where 
\bea
\Delta f(E;\Fermi ,U_0) & =f(E;\Fermi +U_0/2)  -  f(E;\Fermi -U_0/2 ),\nonumber\\ 
\label{eq:DeltaT}
\Tchar  & = T(E;+U_0)-T(E;-U_0), \\
\Tspini &  =T_{\mathrm{S}}(E;+U_0)-T_{\mathrm{S}}(E;-U_0).\nonumber
\eea
%
%
An extension to an adiabatic harmonic driving is straight forward. 

In linear response, the ratchet spin current $\langle I_{\mathrm{S}}\rangle$,  Eq.~\BZ{Isnet},
vanishes because $\Delta T_\mathrm{S}(E;U_0=0)=0$, see Eq.~\BZ{eq:DeltaT}. Hence we must consider 
the nonlinear regime to obtain a finite net spin current. 
Since we consider nonlinear transport ignoring inelastic processes, we can write 
the currents Eq. (2.4) and (2.6) as energy integrals over the transmission.
To model a finite voltage drop across the two leads, we add the term 
\be
\hat{\mathcal{H}}_{U}=U g(x,y;U) 
\label{eq:HU}
\ee
to the Hamiltonian (\ref{Hsim}), where for the square wave driving considered here, 
$U$ takes the values $\pm U_0$ respectively. Here the function $g(x,y;U)$ describes the 
spatial distribution of the electrostatic potential inside the mesoscopic system and is generally obtained through 
a self-consistent solution of the many-particle Schr\"odinger equation and the Poisson equation~\cite{buettiker:1993,christen:1996}.
However, here we employ heuristic models for $g(x,y;U)$, assuming that the voltage primarily
drops in regions, where the magnetic field strongly varies spatially. 
This model is based on the fact that the corresponding Zeeman term acts as an effective 
potential barrier, and takes into account that a more rapid potential variation leads to 
enhanced wave reflection and hence to a steeper local voltage drop~\cite{xu:1993}
(details will be given in the following sections).\\
In order to numerically evaluate the transport properties of the system, the stationary Schr\"odinger equation 
$(\Ham -E)\Phi(\vec{r})=0$ is discretized on a square lattice, yielding a tight-binding representation of $\Ham$. 
This is then used to calculate the elements of the scattering matrix of the system via lattice Greens functions and 
a recursive Greens function algorithm~\cite{lassl:2007}. 


\section{Setup A: Two ferromagnetic stripes with longitudinal magnitization}\label{S3}
In this section we investigate the spin-dependent transport properties (and thereby also the operability as a ratchet) of a quantum wire in the plane of the 2DEG subject to the magnetic field  of two 
stripes with opposite, longitudinal in-plane magnetizations, $\vec{M}=\pm M\hat{y}$, 
arranged perpendicular to the wire. This setup A is shown in Fig.~\ref{figure:1}a). 
The Hamiltonian of the system is given by Eq.~(\ref{Hsim}) with $A_y=0$ (in the Coulomb gauge),
\emph{i.e.} $\Pi_y = p_y$.
For the following analysis we chose a confinement potential $V(y)$ such that 
the wire of width $W$ is displaced by a shift $y_0$ with respect to the symmetric 
configuration, see Fig.~\ref{figure:1}.

For sufficiently narrow wires $(W\ll b)$ and small displacement ($y_0\ll b$), the energy scales of the magnetic field
contributions to $\Ham _0$ in Eq.~(\ref{Hsim}) containing $\vec{B}$ and $\vec{A}$ 
are much smaller than differences between energy levels $E_m-E_n$ of different
transversal modes $|m \rangle $ and $| n \rangle $. Therefore transitions between
different transversal modes are strongly  suppressed, and we consider
the case of only one open mode. Higher modes $|n \rangle$  mimic the
behavior of the first mode up to an energy offset $E_n-E_1$.

Evaluating Eq.~(\ref{Bfin}) for the combined magnetic field of two stripes centered 
around $(x=0,y=0)$, setup A, we obtain the following symmetry properties for the
$B$-field components of this configuration:
%
%
\numparts
\be
B_x(x,-y,z)=-B_x(x,y,z),
\label{Bxsym}
\ee
\be
B_y(x,-y,z)=B_y(x,y,z),
\label{Bysym}
\ee
\be
B_z(x,-y,z)=-B_z(x,y,z) \, .
\label{Bzsym}
\ee
\endnumparts
%
%
%
In particular, Eqs.~(\ref{Bxsym}) and (\ref{Bzsym}) imply vanishing magnetic 
field components $B_x$ and $B_z$ for $y=0$. 
Therefore, we use $\hat{y}$ as the spin quantization axis for the considerations below. 
Those symmetries also have an important implication for the spin dynamics of the system. 
For a confinement potential that is symmetric upon reflection at the $(x,z)$-plane,
$V(-y)=V(y)$, as realized for a symmetric confinement centered around $y_0=0$, 
spin-up and spin-down eigenstates within the same transversal mode $n$ are decoupled.
The relevant matrix element $\bra{n,\sigma}\vec{\sigma}\cdot\vec{B}(x,y)
\ket{n,-\sigma}$, responsible for the spin mixing, vanishes,
%
%
\be
\int_{-\infty}^{\infty} \hspace{-0.2cm}\mathrm{d}y |\chi_n(y)|^2 \Big[ B_z(x,y)-i\sigma
B_x(x,y)\Big]=0 \, ,
\label{perMatel}
\ee
%
%
since the integrand is an odd function of $y$ due to Eqs.~(\ref{Bxsym},\ref{Bzsym}) and 
the fact that transversal modes obey $\chi _n(-y)=(-1)^{n-1}\chi _n(y)$. 
However, for finite values of $y_0$, the coupling \BZ{perMatel} does not vanish anymore, 
and spin flips can arise with a significant effect on electron transport, as we will 
demonstrate in the next paragraph.
%
%
%
%
\subsection{DC transport}
\begin{figure}[tb]
  \begin{center}
	\includegraphics[width=4in]{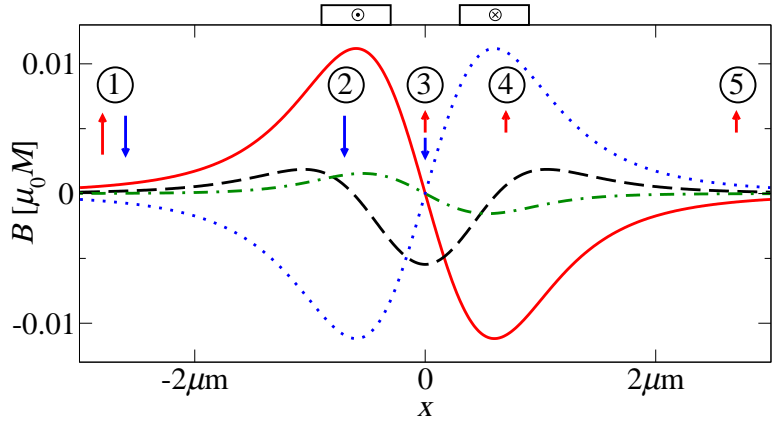}
\caption{\label{figure:2} Magnetic field components $B_x$ (dash-dotted green line), 
$B_y$ (solid red), $-B_y$ (dotted blue) and $B_z$ (dashed black) in the plane of the 
2DEG at fixed $y=200$nm for setup A, Fig.~\ref{figure:1}(a) and parameters given
in the text.}
\end{center}
\end{figure}

Before evaluating the average charge current \BZ{Icnet} and spin ratchet current 
\BZ{Isnet} it is instructive to analyze the DC transport properties of setup A. To 
this end we chose as realistic parameters for the geometry (see Fig.~\ref{figure:1}(a)) 
$W =$ 120 nm, $a=600$ nm, $b=2 \mu$m (thus $W/b=0.06$), $c=200$ nm, $d=600$ nm and $e=100$ nm. 
For this parameter set and $y_0\ll b$ the magnetic field component $B_y(x,y)$ is 
approximately constant in $y$-direction, \emph{i.e.} $B_y(x,y)\approx B_y(x)$. It possesses a 
much larger maximum value than the other components $B_x$ and $B_z$. 
In Fig.~\ref{figure:2} we show the $x$-dependence of the overall magnetic fringe field 
of the two-stripe setup A for fixed $y=200$nm.
%
%
%
%

\begin{figure}[tb]
\begin{center}
	\includegraphics[width=3in]{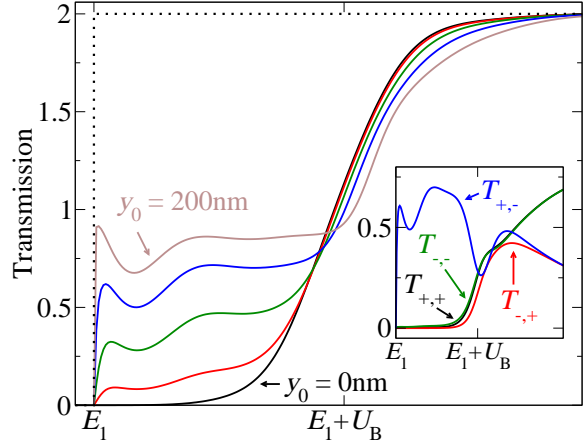}
	\caption{\label{figure:3}Total transmission $T(E)$ in linear response 
for values of $y_0=0$ nm to 200 nm in steps of 50 nm from bottom (black line) 
to top (brown line). The dotted black line 
indicates the transmission for $\vec{B}=0$. Inset: 
spin-resolved transmission probabilities $T_{\sigma ',\sigma}$, Eq.~(\ref{TransDef}),
for $y_0=100$ nm.}
\end{center}
\end{figure}

Figure~\ref{figure:3} shows the total transmission $T(E)$ in linear response ($U_0\rightarrow 0$) for energies within 
the first transversal subband for different values of $y_0$. 
For $y_0=0$ the spin eigenstates decouple owing to Eq.~\BZ{perMatel}. The
energy where the first transversal mode opens is shifted to $E\approx E_1+\UB$ 
due to the Zeeman barrier of height \mbox{$\UB =g^*\mu _{\mathrm{B}}\max [B_y(x,y\!=\!0)]/2$} 
present in the wire. However, for increasing $y_0$ when spin flips can take place,
an additional plateau builds up at energies $E_1 \leq E \leq E_1+\UB$ approaching 
$T\approx 1$. In the inset of Fig.~\ref{figure:3} the spin-resolved 
transmission probabilities $T_{\sigma ',\sigma}$ are depicted for $y_0=100$ nm.
We identify $T_{+,-}(E)$ as the sole contribution to the total transmission 
$T(E)$ for $E_1 \leq E \leq  E_1+\UB$. Thus the appearance of the 
additional plateau is a consequence of the mixing of the spin states. 
For energies well above the barrier both, spin-up and spin-down electrons are fully 
transmitted.
%
%
%
%
 
The main features in the numerically calculated transmission in Fig.~\ref{figure:3} 
can be understood using a heuristic model. 
It is based on the fact that in the region close to $x=0$ (see Fig.~\ref{figure:2})
spin flips predominantly take place, since $B_y(x,y)=0$ vanishes at 
$x=0$, and spin-up and -down states of the same transversal mode are nearly
energy degenerate.

In the following we consider stepwise (positions labeled in Fig.~\ref{figure:2})
the spin evolution along the wire for unpolarized electrons entering the system
with energy $E_1 \leq E \leq E_1+\UB$: 
%
\begin{list}{\textbf{\pscirclebox[linewidth=0.1mm]{\arabic{enumi}}}}{\usecounter{enumi}\setlength{\labelwidth}{2cm}}
	\item Unpolarized electrons (equal number of spin-up and spin-down
	particles) are injected from the left reservoir.
	\item Spin-up electrons are completely reflected at the Zeeman barrier
	(indicated by the solid red line), while spin-down electrons experience a 
	potential valley (blue dotted line) and can pass.
	\item A fraction of the spins is flipped from down to up due to a finite 
	$B_z(x)$ close to $x=0$. 
	\item Spin-down electrons are completely reflected while spin-up electrons pass.
	\item Only spin-down electrons from the left lead reach the right lead, 
	after undergoing a spin flip.
\end{list}
%
%
Hence this mechanism leads to $T_{+,+}=T_{-,+}=T_{-,-}=0$ and $T_{+,-}\neq 0.$
Although the model can explain the basic features of the transmission curves shown 
in Fig.~\ref{figure:3} fairly well, it cannot account for the details in the
functional dependence of $T(E)$ for energies below the barrier which reflects further 
quantum effects present, \emph{e.g.}~resonant tunneling processes.
An analysis of the spin-resolved transmission probabilities in the opposite rocking 
situation, $\mu _{\mathrm{L}}<\mu _{\mathrm{R}}$, where electrons flow from 
right to left, shows that transmitted particles are oppositely spin-polarized compared
with the former case. Correspondingly, $T'_{-,+}$ is the only non-zero component of the 
spin-resolved transmission for energies $E \leq  E\leq E_1+\UB$.\\
The above analysis demonstrates that the magnetic field components 
perpendicular to the dominant one, even if they are small, can significantly alter the 
transport properties of the system. In the present case, disregarding $B_x$ and $B_z$ 
would have resulted in a vanishing transmission for $E<E_1+U_\mathrm{B}$.
%
%
\subsection{AC transport}
We now investigate the rectification properties of setup~A upon applying
the AC driving given in Eq.~(\ref{adrocking}). We first specify how we obtain
the drop of the electrostatic potential $g(x,y;U)$ across the system
for a finite applied bias. 
Based on a heuristic model used in Ref.~\cite{linke:2002}
we assume that $(\partial g(x)/\partial x) 
\propto |(\partial/\partial x)B_y(x,y=0)|$ 
in the central scattering region, $-L/2<x<L/2$, yielding
%
%
\be
g(x)=\frac{1}{2}-\frac{\int_{-L/2}^{x}\mathrm{d}x \left|(\partial /\partial x) B_y(x,y=0) \right|}
{\int_{-L/2}^{L/2}\mathrm{d}x \left|(\partial /\partial x) B_y(x,y=0) \right|},
\label{Eq:2}
\ee 
%
while we fix $g(x)$ to $\pm 1/2$ inside the left (right) lead.
For the full system Hamiltonian $\Ham =\Ham _0+\Ham _U$, Eqs.~(\ref{Hsim},\ref{eq:HU}), 
at finite bias $U_0$ we evaluate the expressions (\ref{Icnet},\ref{Isnet}) for the
average charge and spin currents. If orbital effects due to the perpendicular 
magnetic field $B_z$ are negligible, \emph{i.e.}~$A_y\simeq0$, the total Hamiltonia $\Ham$ 
is invariant under the symmetry operation $\hat{\mathcal{P}}=\hat{R}_x\hat{R}_U\sigma _z$.
Then, as shown in~\ref{AppB}, the relation \BZ{symeq2} between $S$-matrix elements
holds true. Squaring $S$-matrix elements in Eq.~\BZ{symeq2} and summing 
over channels $(n\sigma )\in \mathrm{L}$ and $(n'\sigma ')\in \mathrm{R}$ 
yields $T(E;\pm U_0)=T'(E;\mp U_0)$. This relation, together with
the relation $T'(E;\mp U_0)=T(E;\mp U_0)$ due to unitarity of the $S$-matrix, leads to 
$\langle I_\mathrm{C}\rangle=0$. 
A vanishing average charge current is in line
with symmetry considerations for charge ratchets and coincides with
a numerical analysis for the parameters used here.

\begin{figure}[tb]
\begin{center}
	\includegraphics[width=3in]{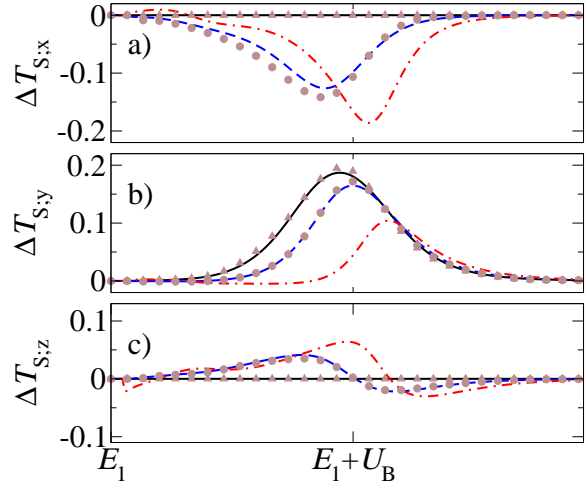}
\caption{\label{figure:4} Ratchet spin transmission $\Delta T_{\mathrm{S}}$ for spin quantization axes $\hat{x}$ (Panel a), $\;\hat{y}$ 
(Panel b) and $\hat{z}$ (Panel c) as a function of the injection energy for displacement
\mbox{$y_0=0$ nm} (solid black line), 100 nm (dashed blue) and 200 nm (dash-dotted red) for 
bias potential $U_0 = 0.1 \UB$ and $g(x)$ specified in Eq.~\BZ{Eq:2}. 
For comparison, $\Delta T_{\mathrm{S}}$ at $y_0=0$ nm (brown triangles) and $y_0=100$ nm 
(brown circles) is shown for a linear voltage drop $\tilde{g}(x)=-x/L$ across
the central scattering region with bias potential $\tilde{U}_0\approx 2.3 U_0$.\\
}
\end{center}
\end{figure}

However, on the other hand, the symmetry considerations imply that the average spin 
current can take finite values. To confirm this numerically and to get an idea of
its magnitude we calculate the ratchet spin current in the right lead according
to Eq.~\BZ{Isnet}. Figure~\ref{figure:4} shows the differences in spin transmissions,
$\Delta T_{\mathrm{S}}(E)$, for the two rocking situations as a function of energy
for moderate finite applied bias voltage $U_0=0.1\UB$. 
For $y_0=0$, where transitions between spin-up and spin-down states of the same 
transversal subband are absent, the system is comparable to the devices studied 
in Ref.~\cite{Proc}. There it was shown, in analogy to the case of charge 
rectification~\cite{Sablikov}, that for conserved spin eigenstates the ratchet effect 
stems from different maximum values $\Delta_\mathrm{max}(U_0)$  of the effective 
potential landscape in the two rocking situations,
\be
\label{eq:Deltamax}
\Delta_\mathrm{max}(U_0)=\max [U_{\mathrm{eff},\sigma}(x,+U_0)]-\max [U_{\mathrm{eff},\sigma}(x,-U_0)] ,
\ee
with 
\be
\label{eq:Ueff}
U_{\mathrm{eff},\sigma}(x,\pm U_0)=\pm U_0g(x)+ (\sigma / 2)
\mu _\mathrm{B}g^* B_y(x,y=0) \, .
\ee
This mechanism explains the rectification related to 
$\Delta T_{\mathrm{S};\hat{y}}(E;y_0=0)$ (solid black line in Fig.~\ref{figure:4}(b)) 
and its functional dependence for $y_0=0$. There spin flips are absent, and 
hence $T_{+,-}\!=\!T_{-,+}\!=\!0$. 
For increasing $y_0$ the magnitude of 
$\Delta T_{\mathrm{S};\hat{y}}(E)$ decreases, while, at the same time,
$\Delta T_{\mathrm{S};\hat{x}}(E)$ and $\Delta T_{\mathrm{S};\hat{z}}(E)$ grow and
take finite values. Thus we can summarize that the ratchet effect survives 
in the presence of mixing of different spin states. However, as apparent 
from Fig.~\ref{figure:4}, the vector of spin polarization is no longer aligned 
along $\hat{y}$ for finite $y_0$.\\
%
%
%
%
%
%
\begin{figure}[tb]
\begin{center}
	\includegraphics[width=3in]{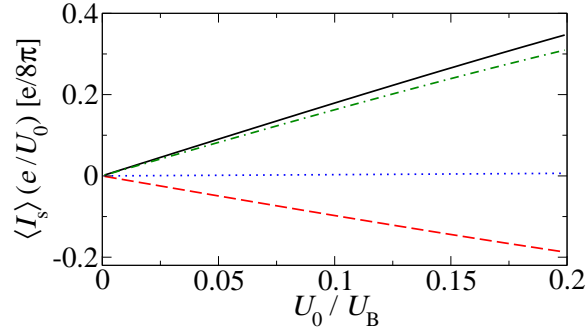}
\caption{\label{figure:5} Bias voltage dependence of the ratchet spin conductance 
$\langle I_{\mathrm{S};\hat{u}}\rangle (e/U_0)$ at zero temperature, $k_\mathrm{B}T=0$, 
for a Fermi energy $\Fermi =E_1+U_B$. Results are shown for $y_0=0$ and polarization
axis $\hat{u}=\hat{y}$ (black solid line), and for $y_0=100$ nm and $\hat{u}=\hat{x}$ 
(red dashed line), $\hat{y}$ (green dash-dotted line) and $\hat{z}$ (blue dotted line).}
\end{center}
\end{figure}
%
%
We further study how sensitively the observed effect depends on the particular 
form of the voltage drop $g(x)$. To this end in Fig.~\ref{figure:4} we additionally 
show $\Delta T_{\mathrm{S}}$ for a linear voltage drop model 
$\tilde{g}(x)=-x/L$ (brown circles/triangles) inside 
the central region $(-L/2\!<\!x\!<\!L/2)$, where the bias voltage $\tilde{U}_0$ was chosen such that 
the maximum value of the respective effective potential (\ref{eq:Ueff}) 
was the same for both voltage drop models:
%
$ \max [U_{\mathrm{eff},\sigma}(x,\pm U_0)]=
\max [\tilde{U}_{\mathrm{eff},\sigma}(x,\pm \tilde{U}_0)] $.
%
%
Comparing $\Delta T_{\mathrm{S}}(E;y_0)$ for $\tilde{g}(x)$ at $y_0=0$ 
(brown triangles) and $100$ nm (brown circles) with the respective function
$\Delta T_{\mathrm{S}}(E;y_0)$ for $g(x)$ we observe no 
significant difference in their functional dependence on $E$, 
although the magnitude of the overall bias is different for both models 
($\tilde{U}_0\approx 2.3U_0$ for the curves presented in Fig.~\ref{figure:4}). 
Therefore we can state that $\Delta T_{\mathrm{S}}(E;y_0)$ rather depends on 
the difference (\ref{eq:Deltamax}) in the maximum values of the effective 
potential 
than the actual distribution of the electrostatic potential in the 
mesoscopic conductor. A study of $\Delta T_{\mathrm{S}}$ in the other lead 
shows very similar results to the ones presented in Fig.~\ref{figure:4}.

In Fig.~\ref{figure:5} we finally display the ratchet spin conductance 
$\langle I_{\mathrm{S}}(\Fermi ,U_0)\rangle (e/U_0)$, Eq.~(\ref{Isnet}), which 
shows a nearly linear dependence on the bias voltage $U_0$, \emph{i.e.}  
$\Delta T_{\mathrm{S}}(E;y_0)\propto U_0$. This is in line with the above 
analysis showing that 
$\Delta T_{\mathrm{S}}(E;y_0) \propto \Delta _\mathrm{max}(U_0)$
and $\Delta _\mathrm{max}(U_0) \propto U_0$.
%
%
%
\section{Setup B: a single ferromagnetic stripe with transverse magnetization}
\label{S4}
\begin{figure}[tb]
\begin{center}
	\includegraphics[width=3in]{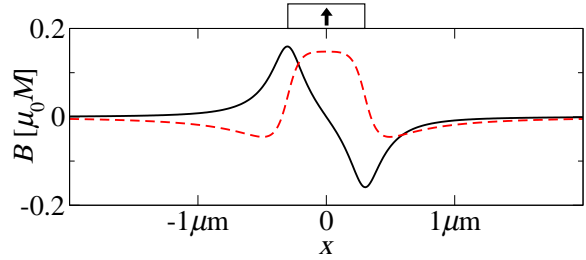}	
\caption{\label{figure:6} Magnetic field components $B_x(x)$ (solid black line) 
and $B_z(x)$ (dashed red line) in the plane of the 2DEG produced by a 
ferromagnetic stripe ($a=600$ nm, $b\rightarrow \infty$, $c=200$ nm, $e=100$ nm,
see Fig.~\ref{figure:1}(b))
with magnetization $\vec{M}=M\hat{z}$.}
\end{center}
\end{figure}

In the following we investigate the possibility of generating a spin ratchet effect 
using the magnetic field profile of a single stripe 
magnetized in the ($x$,$z$)-plane as shown in Fig.~\ref{figure:1}(b).
For $b\gg W$ the evaluation of Eqs.~(\ref{Bfin},\ref{Afin}) for a
magnetization $\vec{M}=M_x\hat{x}+M_z\hat{z}$ yields
%
%
\bdm
\vec{B}(\vec{r})=
\left(\begin{array}{c}
B_x(x,z)\\0\\B_z(x,z)
\end{array}\right),\quad
\vec{A}(\vec{r})=
\left(\begin{array}{c}
0\\A_y(x,z)\\0
\end{array}\right).
\edm
The Hamiltonian of this system then reads
%
\bdm
\hat{\mathcal{H}}_0=\frac{p_x^2+\Pi_y(x)^2}{2m^*}+g^*\frac{\mu_\mathrm{B}}{2}
\Big[ B_x(x)\sigma_x + B_z(x)\sigma_z \Big] +V(y).
\edm
%
The 2DEG is located $e=100$ nm 
below the surface of the semiconductor heterostructure (see Fig~\ref{figure:1}(b)). 
The extension of the stripe in $x$-direction is chosen to be $a=600$ nm, 
infinite in $y$-direction and $c=200$ nm in $z$-direction.
For the analysis below we chose a stripe magnetization $\vec{M}=M\hat{z}$. 
The corresponding magnetic field in the plane of the 2DEG is 
depicted in Fig.~\ref{figure:6}. 
Results comparable to those presented below are obtained for a stripe magnetized 
in $x$-direction. 
%
%
%
%
%
%
%
\subsection{DC transport}\label{K51STra}
%
%
%
\begin{figure}[tb]
\begin{center}
	\includegraphics[width=3in]{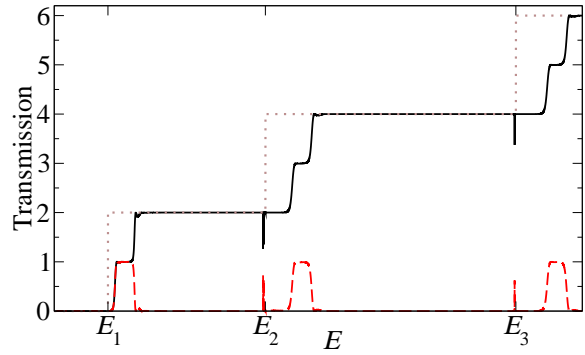}
\caption{\label{figure:7}Total transmission $T(E)$ (solid black line) and absolute 
value of the spin transmission $|T_{\mathrm{S}}|$ (dashed red line) as a function of energy for a wire 
underneath a single ferromagnetic stripe, see Fig.~\ref{figure:1}b and text. For
comparison, the dotted brown staircase function shows the transmission
in the absence of a magnetic field.}
\end{center}
\end{figure}
%
%
%
In Fig.~\ref{figure:7} we show the total transmission $T(E)$ for a quantum wire
of width \mbox{$W=120\mathrm{nm}$} subject to the $B$-field, Fig.~\ref{figure:6}, 
in linear response ($U_0\rightarrow 0$). In addition to the steps at even values of 
$T(E)\approx 2,4,6...$ due to the successive opening 
of the transversal modes at energies $E_n=[\hbar ^2 \pi ^2/ (2m^*W^2)] n^2$,
additional plateaus appear at odd values of $T(E)\approx 1,3,5...$ close to 
the energies $E_n$.
As in Section~\ref{S3}, we can attribute these features to the 
lifted spin degeneracy due to the Zeeman field, since also 
the width of these plateaus corresponds to twice the absolute height 
of the Zeeman barrier $U_{\mathrm{B}}=(g^*/2)\mu _{\mathrm{B}}\max [\, |\vec{B}| \, ]$ 
inside the wire.

In Fig.~\ref{figure:7} we furthermore plot the absolute value of the spin 
transmission,
$|T_{\mathrm{S}}|=\sqrt{(T_{\mathrm{S};x})^2+(T_{\mathrm{S};y})^2+(T_{\mathrm{S};z})^2}$, 
which approaches unity at energies of the additional plateaus.  A
closer look at the spin- and mode-resolved transmission probabilities reveals
that the transmission of the highest occupied transversal subband is completely 
spin polarized at the plateaus, while the lower modes are fully transmitting 
spin-up and spin-down particles. Similar results were reported in 
Ref.~\cite{FengZhaiStripe} for a stripe 
magnetized in the $x$-direction.

Apart from the spin effects due to the Zeeman term, the  
vector potential component $A_y$, affecting the orbital dynamics of the 
electrons due to the perpendicular magnetic field $B_z$, influences the 
electron transport.  In a classical picture,  $B_z$ forces electrons to
move on segments of cyclotron orbits in the plane of the 2DEG.  Therefore, 
the kinetic energy in the direction of motion is reduced resulting in a shift
of the energies where the transversal modes open towards higher 
values~\cite{GovBoese}. This is visible in Fig.~\ref{figure:7}, when comparing 
the total transmission with (solid black line) and without (dotted brown line) 
magnetic field.

\subsection{AC transport}
As for setup A we employ a heuristic model for the voltage drop inside the mesoscopic conductor, 
assuming $\partial g(x) / \partial x \propto |(\partial / \partial x )
|\vec{B}(x)||$ inside the central region $(-L/2 \! < \!x\!<\! L/2)$ yielding
%
%
%
\bdm
g(x)=\frac{1}{2}-\frac{\int_{-L/2}^{x}\mathrm{d}x \left|(\partial /\partial x) |\vec{B}(x)|\right|}
{\int_{-L/2}^{L/2}\mathrm{d}x \left|(\partial /\partial x) |\vec{B}(x)|\right|} \, .
\edm
%
%
%
However, before we numerically investigate the AC ratchet transport properties, 
we exploit certain symmetries present in the system to simplify the expressions 
for the average net charge~\BZ{Icnet} and spin currents~\BZ{Isnet}. 
For the magnetic field profile produced by a stripe magnetized in $z$-direction it is straightforward to 
show from Eqs.~(\ref{Bfin},\ref{Afin}) that the following symmetry relations hold true (see also Fig.~\ref{figure:6}):
%
%
%
\bea
B_x(-x)=-B_x(x) \quad ,&\qquad&
B_z(-x)=B_z(x) \, ,\nonumber\\
A_y(-x)=-A_y(x) \quad ,&\qquad&g(-x)=-g(x) \, .\nonumber
\eea
%
%
%
%
\begin{figure}[tb]
	\begin{center}
		\includegraphics[width=3in]{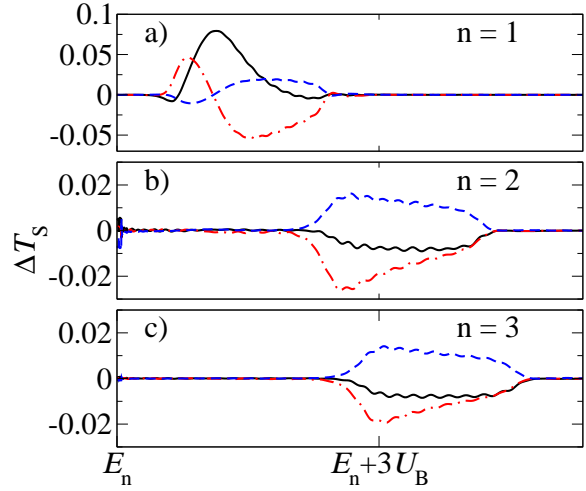}
	\caption{\label{figure:8}Averaged net spin transmission 
		$\Delta T_{\mathrm{S}}(E;U_0)$ for spin polarization axis $\hat{x}$ (solid black line), $\hat{y}$ 
		(dash-dotted red line) and $\hat{z}$ (dashed blue line) as a function of
		energy close to the lowest three transversal energy levels 
		$E_n$ $(n=1,2,3)$ for an applied bias voltage of $U_0=0.1U_\mathrm{B}$.}
	\end{center}
\end{figure}
%
%
Thus the Hamiltonian $\Ham=\Ham _0+\Ham _U$ is invariant under the action of the operator 
$\hat{\mathcal{P}}=-\mathrm{i}\hat{\mathcal{C}}\hat{R}_x\hat{R}_U\sigma _z$, 
where $\hat{\mathcal{C}}$ is the operator of complex conjugation, $\hat{R}_x$ inverses 
the $x$-coordinate, $\hat{R}_U$ changes the 
sign of the applied voltage $(\pm U_0 \leftrightarrow \mp U_0)$ and $\sigma _z$ is the Pauli spin operator.
Due to $[\Ham, \hat{\mathcal{P}}]=0$ we are able to interrelate the transmission 
probabilities for both rocking situations as shown in \ref{AppB}.
Taking the square of Eq.~(\ref{equation_3b}) and summing over the transversal modes 
$n\in \mathrm{L}$ and $n'\in \mathrm{R}$ we obtain the following relations between 
the spin-resolved transmission probabilities in the two rocking situations:
%
%
%
\bdm
T^{(\theta,\phi)}_{\sigma ,\sigma '}(E,\pm U_0)=T^{(\theta,-\phi +\pi)}_{\sigma ',\sigma }(E,\mp U_0).
\edm
%
%
Here the superscript labels the angles of the spin quantization axis on the Bloch sphere (see \ref{AppB}). 
Thus the ratchet charge current \BZ{Icnet} vanishes,
\bdm
\langle I_{\mathrm{C}}(\varepsilon _\mathrm{F} ,U_0)\rangle = 0 \, ,
\edm
and we can express the ratchet spin current~\BZ{Isnet} through the transmission 
probabilities of a single rocking situation (\emph{e.g.} $+U_0$):
\begin{eqnarray}
\langle I_{\mathrm{S};x}(\varepsilon _\mathrm{F} ,U_0)\rangle &=&\frac{1}{4\pi}\int_{0}^{\infty} \mathrm{d}E \;\Delta f(E;\varepsilon _\mathrm{F} ,U_0)\;\times\nonumber\\
&\times&\Big[T_{+,+}(E,+U_0)-T_{-,-}(E,+U_0)\Big] , \nonumber \\
\langle I_{\mathrm{S};y/z}(\varepsilon _\mathrm{F} ,U_0)\rangle &=&\frac{1}{4\pi}\int_{0}^{\infty} \mathrm{d}E \;\Delta f(E;\varepsilon _\mathrm{F} ,U_0)\;\times\nonumber\\
&\times&\Big[T_{+,-}(E,+U_0)-T_{-,+}(E,+U_0)\Big] \,  . \nonumber
\end{eqnarray}
%
%
%
%
Figure~\ref{figure:8} shows the ratchet 
spin transmission $\Delta T_{\mathrm{S}}(E;U_0)$ at a finite applied 
voltage $U_0=0.1U_B$ for a wire of width $W=120$ nm. 
This quantity is finite for energies where the DC transmission is spin polarized 
(see Fig.~\ref{figure:7}). Furthermore, the spin polarization of the ratchet spin 
transmission depends on the injection energy. This opens the possibility to tune it 
upon varying the Fermi energy. 
Comparing the ratchet spin transmission for $n\!=\!1$ (Panel a) and 
$n\!=\!2,3$ (Panels b,c) we observe that its magnitude is significantly 
lower in the case where more than one transversal mode is conducting. 
This behavior is due to the mixing of different transversal subbands due 
to $A_y(x)$. To quantify this effect we introduce 
\bdm
\Delta T_\mathrm{S,max}(n)=\max_{E\in[E_n,E_{n+1}]}\!
\left[\sqrt{\Delta T_{\mathrm{S};\hat{x}}(E)^2+\Delta T_{\mathrm{S};\hat{y}}(E)^2+\Delta T_{\mathrm{S};\hat{z}}(E)^2}\right] 
\edm
as a measure for the rectification in each single transversal mode.
Figure~\ref{figure:9} shows that $\Delta T_\mathrm{S,max}(n=2)/\Delta T_\mathrm{S,max}(n=1)\approx 1$ 
for cases where the mixing due to 
$A_y(x)$ is small, \emph{i.e.}~for a narrow wire and/or small magnetic field. However it decreases 
upon increasing $\mu _0M$ and/or $W$. Note that for setup A in the previous section 
the magnetic field inside the quantum wire 
was one order of magnitude smaller than for setup B here, thus 
yielding a comparable value $\Delta T_\mathrm{S,max}(n)$ for all subbands $n$.
%
%
\begin{figure}[tb]
\begin{center}
	\includegraphics[width=3in]{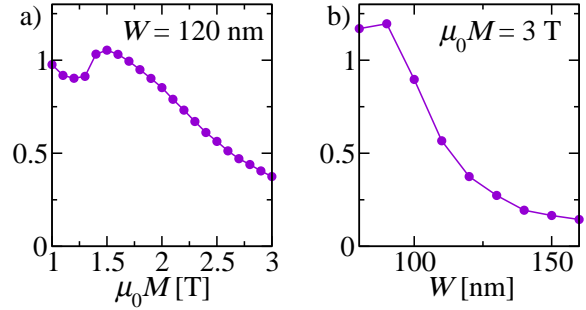}
\caption{\label{figure:9} Ratio $\Delta T_\mathrm{S,max}(n=2)/\Delta T_\mathrm{S,max}(n=1)$ as a function of (a) the magnetization of the ferromagnetic stripe and (b) 
the wire width.}
\end{center}
\end{figure}
%
%

\begin{figure}[tb]
\begin{center}
	\includegraphics[width=3in]{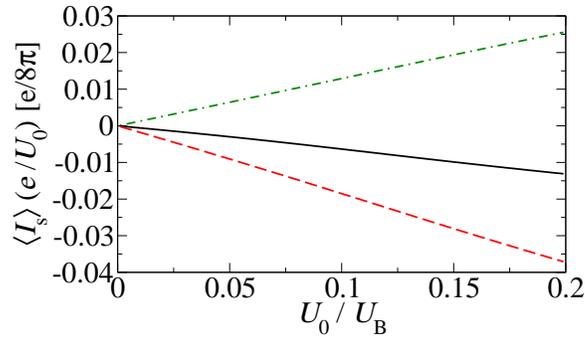}
\caption{\label{figure:10}Ratchet spin conductance 
$\langle I_{\mathrm{S},\hat{u}}\rangle (e/U_0)$ as a function of bias voltage 
$U_0$ at  $k_\mathrm{B}T=0$ and $\Fermi =E_3+3U_B$ for the spin quantization
directions $\hat{x}$ (black solid line), 
$\hat{y}$ (red dashed line) and $\hat{z}$ (green dash-dotted line).}
\end{center}
\end{figure}

As for setup A, the ratchet spin conductance shown in Fig.~\ref{figure:10}
exhibits a linear dependence on the applied voltage. Thus we presume that 
the rectification mechanism is the same as in setup A, although the spin dynamics 
is much more intricate. 
%
%
%
%
\section{Conclusions and Outlook}\label{Conclusions}
In the present work we have shown that the coupling of the electron spin to the magnetic fringe fields of ferromagnetic stripes via the Zeeman interaction can be used to generate a spin 
ratchet effect in a coherent mesoscopic conductor subjected to an adiabatic AC driving 
with finite bias. The proposed devices exhibit the appealing property of creating a directed
net spin current in the absence of an accompanying net charge transport.
This key result has been demonstrated in numerical approaches for setups A and B, 
in the case of setup B, also analytically based on symmetry properties of the system.

The generated spin current may be regarded as resulting from a rectification effect,
however in a generalized sense: The direct analogue of a charge current rectifier
would be a system generating a directed spin current out of a conductor with
alternating {\em spin-chemical} potentials in the left and right reservoir. Spin ratchets, 
as the ones considered here, act differently as they convert an AC {\em electrical} bias into 
a net spin current.
 
>From our analysis we have identified the difference in the maximum values of the 
effective Zeeman potentials in the respective rocking situations as responsible 
for the creation of the spin current. It has been shown that this rectification effect 
is almost independent of the actual distribution of the electrostatic potential 
in the biased conductor. Furthermore, the fact that, for the systems considered
here, the difference in the maximum values of the Zeeman potential is crucial for 
the spin currents, implies that the magnitude of the spin current cannot be systematically  
increased upon increasing the number of magnetic stripes, \emph{e.g.} in a periodic
arrangement of stripes. We have checked this also numerically by adding an 
increasing number of stripes.

In the preceding sections we presented results, when evaluating the spin 
current inside the right lead. However, as we noted both systems considered, 
setup A and B, are characterized by interesting symmetry properties. Those can now be used 
to directly extract the respective currents inside the left lead. 
If for setup A the component $A_x$ can be neglected, as it is 
appropriate for the parameters used in Sec.~\ref{S3}, the combined Hamiltonian of the 
system and the leads has been proven to be invariant under the action of the symmetry 
operation $\hat{R}_x\hat{R}_y\hat{R}_U\sigma _z$, while for setup B the Hamiltonian 
is invariant under the action of $\hat{R}_x\hat{R}_U\sigma _z$. These symmetry
properties are reflected in Eqs.~\BZ{symeq2} and~\BZ{symeq3} respectively, see \ref{AppB}. 
Both relations lead to the general relation
\bdm
T^{\prime(\theta,\phi +\pi)}_{\sigma ,\sigma '}(E,\mp U_0)=
T^{(\theta,\phi)}_{\sigma ,\sigma '}(E,\pm U_0)
\edm
between the transmission probabilities in the two rocking situations.
%
%
%
This relation allows for the following interpretation. If the transmitted electrons 
are spin-polarized in one of the two rocking situations with direction of
the polarization vector given by the angles $(\theta ,\phi)$ on the Bloch sphere, 
then in the other rocking situation the spin polarization vector of the output current 
(in the other lead) will be rotated around the $z$-axis by $\pi$ and thus points to 
$(\theta ,\phi +\pi)$. 
This property is a direct consequence of the lacking conservation of the spin 
eigenstates inside the wire.\\
The physics of semiconducting materials characterized by a large $g^*$ factor is dominated 
by the presence of magnetic impurities, \emph{e.g.} diluted magnetic 
semiconductors~\cite{FurdDMSRev}. Therefore, in order to exploit the expected 
stronger rectification effect, we have to take elastic scattering off impurities into account. 
In particular, we plan to study how additional disorder alters the spin ratchet effects. 

Finally, since the heuristic model used for the distribution of the electrostatic potential 
in the conductor is convenient but not fully satisfactory, it would be desirable to 
calculate the charge density and the respective electrostatic potential inside the 
wire self-consistently. Work in this direction is in progress.
%
%
%
%
\ack
We acknowledge discussions with and computational support by Michael Wimmer. 
This research has been supported by the Deutsche Forschungsgemeinschaft within the cooperative 
research center SFB 689 ``Spin phenomena in reduced dimensions''. MS acknowledges support
through the Studienstiftung des Deutschen Volkes.

\begin{appendix}
\section{Calculation of the spin current in the Landauer-B\"uttiker formalism}\label{App}
In this appendix we present a derivation of the expressions for spin 
current in the leads of a multiterminal 
coherent conductor within the framework of the Landauer-B\"uttiker theory~\cite{Buttiker}. 
To this  end we consider $N$ non-ferromagnetic contacts, injecting spin-unpolarized 
current into the leads. For convenience we use a local coordinate system for the lead 
under investigation, where $x$ is the coordinate along the lead in the direction of 
charge propagation due to an applied bias in linear response and $y$ is the transverse 
coordinate. 
Then the eigenfunctions inside a lead are given by
%
\be
\Phi^{\pm} _{E,n\sigma}(x,y)=\frac{1}{\sqrt{k_x(E)}}e^{\pm \im k_x(E) x}\chi _n(y)\Sigma (\sigma )
\label{leadeigenfunct}
\ee
%
where the $\chi _n(y)$ are the transverse eigenfunctions of the lead with the transversal eigenenergy 
$E_n$ and $\Sigma (\sigma )$ is the spin eigenfunction. The 
superscript~$\pm$ of $\Phi$ refers to the direction of motion in $\pm x$-direction with the 
wave-vector $k_x=\sqrt{2m^*(E-E_n)}/\hbar$. 
For the derivation we use the scattering approach, where the amplitudes of the states inside the leads are 
related via the scattering matrix $\mathbf{S}(E)$, determined by the Hamiltonian of the coherent 
conductor. 
Inside lead $q$ a given scattering state 
%
\bdm
\varphi ^q_E(x,y)=\!\!\!\!\sum_{(n\sigma )\in q}\!\!\!\left(a^{q}\nsi (E)\Phi^+_{E,n\sigma}(x,y) + 
b^{q}\nsi (E)\Phi^-_{E,n\sigma}(x,y)\right),
\edm
%
%
%
$(\sigma = \pm$),
consists of incoming states $\Phi^+$ entering the coherent conductor from contact $q$ and 
outgoing states $\Phi^-$ leaving the coherent conductor into contact $q$. 
The amplitudes of incoming $a^j_{n\sigma}$ and outgoing waves $b^i_{\nsp}$ are related via the equation 
%
%
%
\begin{equation}
b^i_{n'\sigma '}(E)=\sum^{N}_{j=1}\sum_{n \in j}\sum_{\sigma =\pm 1}
S^{i,j}_{n^\prime \sigma^\prime ,n\sigma}(E)a^j_{n\sigma}(E),
\label{S_matrix_action_definition}
\end{equation}
%
%
%
where the scattering matrix $\mathbf{S}(E)$ has the following structure for an $N$ terminal system:
%
\bdm
{\bf S}(E)=
\left(
\begin{array}{cccc}
{\bf r}^{1,1}(E)&{\bf t}^{1,2}(E)&\cdots&{\bf t}^{1,N}(E)\\
{\bf t}^{2,1}(E)&{\bf r}^{2,2}(E)&\cdots&{\bf t}^{2,N}(E)\\
\vdots          &   \vdots        &\ddots&\vdots      \\
{\bf t}^{N,1}(E)&{\bf t}^{N,2}(E)&\cdots&{\bf r}^{N,N}(E)
\end{array}
\right).
\edm
%
%
%
Here the sub-matrix ${\bf r}^{j,j}(E)$ is a square matrix of dimensionality $M^j(E)$, corresponding 
to the number of open channels at energy $E$ in lead $j$ (already including the spin degree of freedom), which 
is connected to a reservoir with chemical potential $\mu _j$. The matrix ${\bf r}^{j,j}(E)$ contains 
the scattering amplitudes of incoming channels of lead $j$ being reflected back into 
outgoing channels of the same lead.
The sub-matrix ${\bf t}^{i,j}(E)$ is a $M^i(E)\times M^j(E)$ matrix that contains 
the scattering amplitudes for transmission between incoming channels from lead $j$ and 
outgoing channels of lead $i$.\\

The wave function of the scattering state inside lead $i$, where only the incoming channel $(n\sigma )\in j$ is populated 
$(a^{j'}_{n'\sigma '}=\delta _{j',j}\delta _{n',n}\delta _{\sigma ',\sigma})$,  
reads for $j=i$:
%
%
\be
\varphi ^i_{E,n\sigma}(x,y)=
\Phi ^+_{E,n\sigma}(x,y) +\sum_{(n'\sigma ')\in i}r^{i,i}_{\nsp ,n\sigma} (E)
\Phi ^-_{E,\nsp}(x,y),
\label{Mscattstatp}
\ee
and, correspondingly, for $j \neq i$
\be
\varphi ^i_{E,n\sigma }(x,y)=
\sum_{(n'\sigma ')\in i}t^{i,j}_{n'\sigma ',n\sigma } (E) \Phi^-_{E,n'\sigma '}(x,y).
\label{Mscattstatq}
\ee
%
%
For a wave function $\Psi(x,y)$ the spin current $I_{\mathrm{S}}^{\Psi}(x)$ passing a cross 
section ($x=\,$const) of a lead is given by:
%
%
%
\be
I_{\mathrm{S}}^{\Psi}(x)=\int \mathrm{d}y \Psi ^*(x,y)\hat{J}_{\mathrm{S}}\Psi(x,y).
\label{A:Eq:5}
\ee
%
%
%
%
Here we use the most common definition of the spin current operator~\cite{RashbaSpinDyn}, 
which with respect to an arbitrary quantization axis $\hat{u}$ takes 
the following form inside the leads:
%
%
%
\bdm
\hat{J}_{\mathrm{S}}=\frac{\hbar}{2}\frac{\hbar}{2m^*i}\left(\vec{\sigma}\cdot\hat{u}\right)\left( \overrightarrow{\frac{\partial}{\partial x}} 
- \overleftarrow{\frac{\partial}{\partial x}} \right) \, .
\edm
%
The partial derivatives act on the expressions to their right/left respectively (indicated by the arrows). 
For the scattering state (\ref{Mscattstatp}) 
we then obtain for the spin current~\BZ{A:Eq:5} inside lead $i$
%
%
\bdm
I^{j = i}_{\mathrm{S};E,n\sigma}(x\in i)=\frac{\hbar ^2}{2m^*}\left[ \sigma -
\sum_{(n'\sigma ')\in i}\sigma '\left|r^{i,i}_{n'\sigma ',n\sigma}(E)\right| ^2 
\right] \, ,
\edm
where $(n\sigma \in j, j = i)$. For
the scattering state (\ref{Mscattstatq}) we find the corresponding expression ($n\sigma \in j,
j \neq i$)
\bdm
I^{j\neq i}_{\mathrm{S};E,n\sigma}(x\in i)=-\frac{\hbar ^2}{2m^*}\sum_{(n'\sigma ')\in i}
\sigma '\left|t^{i,j}_{n'\sigma ',n\sigma }(E)\right| ^2 \, .
\edm
%
%
Since every channel is populated according to the Fermi-Dirac distribution 
$f(E;\mu _q)$ of the respective contact $q$, the total spin current in lead $i$ 
then reads
%
%
\bea
\label{eq:I_S}
I_{\mathrm{S}}(x\in i) 
& = & \frac{m^*}{2\pi\hbar ^2}\int_{0}^{\infty} \mathrm{d}E \; \Bigg[ 
\sum_{j=1}^{N}\sum_{(n\sigma )\in j}f(E;\mu _j ) I^{j}_{\mathrm{S};E,n\sigma}(x\in i)
\Bigg] \\
& = & \frac{-1}{4\pi}\int_{0}^{\infty} \!\!\!\mathrm{d}E \Big[ f(E;\mu _i)R^{i,i}_{\mathrm{S}}(E)+\sum_{q\neq i}f(E;\mu _q ) 
T^{i,q}_{\mathrm{S}}(E)\Big] \nonumber
\eea
%
where 
%
\bea
T^{i,q}_{\mathrm{S}}(E)& = & \sum_{\sigma '=\pm}\left( T^{i,q}_{+,\sigma '}-T^{i,q}_{-,\sigma '}\right)\nonumber\\
R^{i,i}_{\mathrm{S}}(E)& = & \sum_{\sigma '=\pm}\left( R^{i,i}_{+,\sigma '}-R^{i,i}_{-,\sigma '}\right)\,  ,\nonumber
\eea
%
and
%
\bea\label{TransDef}
T^{i,q}_{\sigma ,\sigma '}(E) & = & \sum_{n\in i}\sum_{n'\in q}\left|t^{i,q}_{n \sigma ,\nsp}(E)\right| ^2 \, ,\\
R^{i,i}_{\sigma ,\sigma '}(E) & = & \sum_{n\in i}\sum_{n'\in i}\left|r^{i,i}_{n \sigma ,\nsp}(E)\right| ^2 \, .
\eea
%
%
Since $\mathbf{S}(E)$ is unitary, the following relation holds true:
%
%
%
%
\bdm
\sum_{(n'\sigma ')\in i}\!\!\left|r^{i,i}_{n\sigma ,\nsp}(E)\right| ^2 +
\sum_{q\neq i}\sum_{(n''\sigma '')\in q}\!\!\left|t^{i,q}_{n\sigma ,n''\sigma ''}(E)\right| ^2 = 1.
\edm
%
%
Then it is straightforward to show that 
%
%
%
\bdm
R^{i,i}_{\mathrm{S}}(E)+\sum_{q\neq i}T^{i,q}_{\mathrm{S}}(E)=0,
\edm
%
In view of Eq.~(\ref{eq:I_S}) we eventually find for the 
spin current in lead $i$
%
%
\be
I_{\mathrm{S}}(x\in\! i)=\!
\frac{1}{4\pi}\int_{0}^{\infty}\!\! \mathrm{d}E 
\sum_{q\neq i}  
\left[ f(E;\mu _i)\!-\! f(E;\mu _q ) \right]
T^{i,q}_{\mathrm{S}}(E) .
\label{spincurr}
\ee
%
%
%
%
%
%
%
Although equilibrium spin currents can locally exist in systems with SO interactions 
as shown for 2DEGs~\cite{Rashba:2003} and in mesoscopic systems~\cite{Nikolic:2006}, 
Equation~\BZ{spincurr} clearly shows that in thermal equilibrium ($\mu _j=\mu$) 
the spin current inside leads without SO interactions and magnetic fields vanishes.
This absence of equilibrium spin currents in the leads has been shown for systems with 
preserved time-reversal symmetry~\cite{NoESCSO}. However, here we show that it is 
more generally valid for any coherent conductor, since we did not make any assumptions 
about symmetries of the scattering region in the course of the derivation.

We note that the relation (5) derived in Ref.~\cite{Pareek} that 
allows for equilibrium spin currents, has to be regarded as incorrect, arising 
from an improper treatment of the back-reflection, missing the term 
%
\be
\label{missing}
\frac{1}{4\pi}\int^{\infty}_{0}\!\!\!\!\!\mathrm{d}E\; f(E;\mu _i) 2
\left( R^{i,i}_{-,+}(E) - R^{i,i}_{+,-}(E) \right)
\ee
%
%
in comparison with Eq.~\BZ{spincurr}. This issue has been addressed by 
Nikolic \emph{et al.}~\cite{Nikofalsch}. 
However, their argumentation that the term~(\ref{missing}) has only to be included in equilibrium, \emph{i.e.} 
for energies up to the lowest chemical potential of the $N$ terminals, seems
questionable. If the spin current is evaluated 
in a lead connected to a reservoir $k$ with $\mu _k>\mu _l$ ($\mu _l$ being the lowest chemical potential 
of any of the $N$ reservoirs), the full expression~\BZ{spincurr} has to be used. 
Furthermore, a simplified version of Eq.~\BZ{spincurr},
\bdm
I_\mathrm{S}=\frac{1}{4\pi}\sum_{q\neq i} (\mu _i -\mu _q) T_\mathrm{S}^{i,q} \, ,
\edm
where transport at zero temperature and energy-independent transmission probabilities 
were considered, has been used in recent publications on the mesoscopic spin Hall 
effect~\cite{Correct}.
%

\section{Derivation of symmetry relations for spin dependent Landauer transport at finite bias}\label{AppB}
Here we derive symmetry relations for a two terminal setup as used in this Paper. We generalize related 
expressions from Ref.~\cite{FengZhaiSymm} to finite bias and arbitrary spin quantization axis. 

If the Hamiltonian of the total system of scattering region and leads is invariant under certain symmetry operations 
$\hat{\mathcal{P}}$, we can relate the elements of the scattering 
matrix even in different rocking situations. As an example we derive the symmetry relations stemming from 
the symmetry operator $\hat{\mathcal{P}}=-\mathrm{i}\hat{\mathcal{C}}\hat{R}_x\hat{R}_U\sigma _z$, 
where $\hat{\mathcal{C}}$ is the operator of complex 
conjugation, $\hat{R}_x$ inverts the $x$-coordinate, $\hat{R}_U$ changes the 
sign of the applied voltage $(\pm U_0 \leftrightarrow \mp U_0)$ and $\sigma _z$ is the Pauli spin operator.

Generalizing Eq.~(\ref{leadeigenfunct}) of \ref{App} the eigenfunctions 
inside the leads are given by 
%
%
\be
\Phi^{\pm ,(\theta ,\phi )} _{E,n\sigma}(x,y)=
\frac{1}{\sqrt{k_x(E)}}e^{\pm \im k_x(E) x}\chi _n(y)\Sigma _{(\theta ,\phi)}(\sigma )
\label{LEigen}
\ee
%
%
with the spin eigenstates
%
\bdm
\Sigma _{(\theta ,\phi)}(+)=\arr{\cos\frac{\theta}{2}e^{-i\phi /2}\\\sin\frac{\theta}{2}e^{i\phi /2}},\qquad
\Sigma _{(\theta ,\phi)}(-)=\arr{-\sin\frac{\theta}{2}e^{-i\phi /2}\\\cos\frac{\theta}{2}e^{i\phi /2}},
\edm
%
defined with respect to the quantization axis 
\bdm
\hat{u}=\left( \begin{array}{c} \sin\theta\cos\phi \\ \sin\theta\sin\phi \\ \cos\theta 
\end{array}\right) .
\edm
The effect of $\hat{\mathcal{P}}$ on the eigenstates~\BZ{LEigen} is to change an incoming state in the left lead (L) in one 
rocking situation into an outgoing state of the right lead (R) in the other rocking situation and \emph{vice versa}. 
Furthermore the position of the spin on the Bloch sphere is changed from $(\theta,\phi)$ into $(\theta,-\phi +\pi)$, and 
the amplitude of the state is complex conjugated.\\
On the other hand, the action of $\hat{\mathcal{P}}$ cannot change the scattering-matrix, since the Hamiltonian is invariant under the action 
of $\hat{\mathcal{P}}$. Therefore the following relation holds true:
%
%
%
\be
a^{(\theta,-\phi +\pi)*}_{\bar{n}\sigma}(E,\mp U_0)=\sum_{n^\prime\in(\mathrm{L}\cup\mathrm{R})}\sum_{\sigma^\prime =\pm 1}
S^{(\theta,\phi)}_{n\sigma ,n^\prime \sigma^\prime}(E,\pm U_0)b^{(\theta,-\phi +\pi)*}_{\bar{n}^\prime \sigma^\prime}(E,\mp U_0)
\label{eq:sym1}
\ee
Here, the  mode index $\bar{n}$ is related to mode $n$ 
of the opposite lead by means of the symmetry transformation.
Comparing Eq.~(\ref{eq:sym1}) with the inverse of Eq.~(\ref{S_matrix_action_definition}),
\bdm
a^{(\theta,-\phi +\pi)}_{n\sigma}(E,\mp U_0)=\\
\sum_{n^\prime\in(\mathrm{L}\cup\mathrm{R})}\sum_{\sigma^\prime =\pm 1}
\left[S^{(\theta,-\phi +\pi)}\right]^{-1}_{n\sigma ,n^\prime \sigma^\prime} (E,\mp U_0)
b^{(\theta,-\phi +\pi)}_{n^\prime \sigma^\prime}(E,\mp U_0) \, ,
\edm
%
%
and using the unitarity of the scattering matrix, \mbox{${\displaystyle\mathbf{S}^{-1}=\mathbf{S}^{\dagger}=(\mathbf{S}^{*})^\mathrm{t}}$}, 
we find the symmetry relation for the scattering amplitudes
%
%
%
\begin{equation}\label{equation_3b}
S^{(\theta,-\phi +\pi)}_{n\sigma ,n^\prime \sigma^\prime}(E,\mp U_0)
=S^{(\theta,\phi)}_{\bar{n}^\prime \sigma^\prime ,\bar{n}\sigma}(E,\pm U_0).
\end{equation}
%
Similar relations can be obtained for any other symmetry operator that commutes with the Hamiltonian $\Ham$. 
For the other two operators used in this paper,
$\hat{R}_x\hat{R}_U\sigma _z$ and $\hat{R}_x\hat{R}_y\hat{R}_U\sigma _z$,
the above procedure can be applied accordingly. It yields
\be\label{symeq2}
S^{(\theta,\phi +\pi)}_{n\sigma ,n^\prime\sigma^\prime}(E,\mp U_0)=S^{(\theta,\phi)}_{\bar{n}\sigma ,\bar{n}^\prime\sigma^\prime}(E,\pm U_0)
\ee
for $[\Ham ,\hat{R}_x\hat{R}_U\sigma _z ]=0$ and
\be\label{symeq3}
S^{(\theta,\phi +\pi)}_{n\sigma ,n^\prime\sigma^\prime}(E,\mp U_0)=p_np_{n'}
S^{(\theta,\phi)}_{\bar{n}\sigma ,\bar{n}^\prime\sigma^\prime}(E,\pm U_0)
\ee
for $[\Ham ,\hat{R}_x\hat{R}_y\hat{R}_U\sigma _z ]=0$, where $p_n=(-1)^{n-1}$ is the parity of the eigenfunction $\chi _n(y)$ in Eq.~(\ref{LEigen}).
\end{appendix}

\end{document}